\documentclass[12pt]{article}

\usepackage{times}

 \usepackage{graphicx}
 \usepackage{dcolumn}
 \usepackage{bm}
 \usepackage{amsmath}
 \usepackage{color}
 \usepackage{amssymb}
 \usepackage{mathrsfs}
 \usepackage{MnSymbol}%
 \usepackage{wasysym}%
 \usepackage{gensymb}
 \usepackage{framed,color}
 \usepackage{enumitem}
 \usepackage[normalem]{ulem}

\newenvironment{sciabstract}{%
\begin{quote} \bf}
{\end{quote}}

\title{Observation of Half-Quantum Flux in Unconventional Superconductor $\beta$-Bi$_2$Pd}

\author
{Yufan Li, Xiaoying Xu, C. L. Chien\\
\\
\normalsize{Department of Physics and Astronomy, Johns Hopkins University, Baltimore, MD 21218, USA}\\
}

\date{}

\begin{document} 


\baselineskip24pt


\maketitle


\begin{sciabstract}
  We report the observation of half-integer magnetic flux quantization in mesoscopic rings of superconducting $\beta$-Bi$_2$Pd thin films. 
  The half-quantum fluxoid manifests itself as a $\pi$ phase shift in the quantum oscillation of the critical temperature. 
  This result verifies unconventional superconductivity of $\beta$-Bi$_2$Pd, in accord with the expectation of a topological superconductor. 
  We also discuss the strong indication that $\beta$-Bi$_2$Pd is a spin-triplet superconductor. 
\end{sciabstract}



The condensation of Cooper pairs gives rise to superconductivity \cite{Tinkham_book}. 
A key signature of the electron pairing is the quantization of the magnetic flux through a multiply-connected superconducting body, in discrete units of $\Phi_0 = hc / 2e$. 
Indeed, the observations of the fluxoid quantization served as the first experimental verifications of the BCS theory \cite{Deaver1961,Doll1961,Byers1961}. 
Shortly after the initial magnetometry measurements, Little and Parks further demonstrated the oscillatory feature of the superconducting transition temperature $T_c$, as a result of the periodic free energy of the superconducting state as a function of the applied magnetic flux \cite{Little1962}. 
The minimum of the free energy, or the maximum of the $T_c$, is always achieved when the applied magnetic flux takes $\Phi = n\Phi_0$, where $n$ is an integer number. 
In the following decades, the Little-Parks effect, as a stringent test for the electron pairing, has been observed in numerous superconducting materials \cite{Parks1964,Gammel1990,Sochnikov2010,cai_unconventional_2013}. 
On the other hand, Geshkenbein, Larkin and Barone (GLB) predicted that half magnetic flux quanta may exist in spin-triplet superconductors (SCs) \cite{geshkenbein_vortices_1987}. 
The idea was later extended into the spin-singlet high-$T_c$ SCs \cite{sigrist_paramagnetic_1992}, and realized in YBa$_2$Cu$_3$O$_{7-\delta}$ tricrystals with delicately designed crystalline boundaries \cite{tsuei_pairing_1994,Kirtley1995}, which, in conjuncture with the corner SQUID experiments \cite{wollman_experimental_1993,Wollman1995}, pinpointed the $d$-wave pairing symmetry. 
However, the effect of half-quantum fluxoid (HQF) for polycrystalline spin-triplet SC loops, as in the original proposal, has never been experimentally demonstrated. 
More recently, experimental indications of a different HQF effect has been reported in the spin-triplet SC candidate Sr$_2$RuO$_4$ \cite{jang_observation_2011,yasui_little-parks_2017}. 
This phenomenon is believed to manifest itself as a splitting of the integer-quantization steps, which is energetically unfavorable unless certain special conditions are met \cite{chung_stability_2007}. 
It is not to be confused with the HQF effect proposed by GLB, in which case the half flux quanta of $(n+1/2)\Phi_0$ are energetically preferred over the integer ones of $n\Phi_0$. 

The past decade witnesses the surging interest for the topological superconductors (TSCs), which may host Majorana fermions \cite{alicea_new_2012,kallin_chiral_2016,Sato2017}. 
The TSCs are considered to have a profound connection to the spin-triplet pairing, or its mathematical equivalents \cite{read_paired_2000,Kitaev_2001,Fu2008}. 
Brought back to the spotlight are the triplet pairing hopefuls including Sr$_2$RuO$_4$ and UPt$_3$, both of which have attracted intensive interest for decades, including indications of HQF in SQUID devices comprising Sr$_2$RuO$_4$ \cite{nelson_odd-parity_2004}. 
However, the experimental evidences still fall short for concluding spin-triplet pairing \cite{kallin_chiral_2016,Sato2017,Mackenzie2017}.
The search goes on in relatively new SC materials, including doped topological insulators, noncentrosymmetric SCs \cite{kallin_chiral_2016,Sato2017}, as well as iron-based SCs \cite{Zhang2018}. 

Of particular interest is $\beta$-Bi$_2$Pd with a centrosymmetric tetragonal crystal structure \cite{Imai2012}, which is reported to host spin-polarized topological surface states that coexist with superconductivity \cite{Sakano2015,Iwaya2017}. 
One scanning tunnelling spectroscopy study, in particular, reports observation of Majorana bound states at the center of the vortices in epitaxial thin films \cite{Lv2017}. 
Controversy persists, however, while other tunneling spectroscopy and calorimetric studies in bulk specimen suggest the conventional $s$-wave pairing mechanism \cite{Herrera2015,Che2016,Kacmrcik2016}. 
We performed the Little-Parks experiment on mesoscropic superconducting rings fabricated on textured $\beta$-Bi$_2$Pd thin films. 
The Little-Parks oscillations shift by one half of a period, or a phase of $\pi$, which is the experimental signature of the HQF predicted by GLB \cite{geshkenbein_vortices_1987}. 
It is unequivocally evident that the superconductivity of $\beta$-Bi$_2$Pd originates from unconventional pairing symmetry. 
The observation resonates with the expectation of a TSC. 
Our result strongly suggests that $\beta$-Bi$_2$Pd is a spin-triplet SC. 

The fluxoid $\Phi^{\prime}$ of a superconducting loop is introduced by F. London as $\Phi^{\prime} =\Phi +(4\pi/c) \oint \lambda^2\vec{j_s}\cdot d\vec{s}$, where $c$ is the speed of light, $\lambda$ is the London penetration depth, and $\vec{j_s}$ is the supercurrent density \cite{LondonF}. 
It is shown that $\Phi^{\prime}$ must take quantized values, with integral increments of a fluxoid quantum $\Phi_0$ \cite{LondonF,Deaver1961,Byers1961}. 
The fact that the applied magnetic flux $\Phi$ can take arbitrary values requires $\vec{j_s}$ to compensate it in order to maintain the quantized $\Phi^{\prime}$. 
This leads to the periodic oscillation of the free energy, or $T_c$, in response to the applied magnetic field \cite{Little1962}. 
In ordinary cases, a circulating $\vec{j_s}$ is not required when the external field provides exactly $\Phi=n\Phi_0$. 
Therefore the free energy of the superconducting state is the lowest; as a result, the $T_c$ is the highest. 
The maximum of the free energy and the minimum of $T_c$ occur when $n-\Phi/\Phi_0=\pm\frac{1}{2}$, as depicted in Fig.~1A. 
For unconventional SCs, the superconducting order parameter becomes anisotropic, retaining the symmetry which represents that of the underlying crystal lattices. 
It is possible, as suggested by GLB, that the phase factor of the complex order parameter will shift an additional phase difference of $\pi$ along a path across the boundary of two crystal grains, inflicting a sign change in the corresponding free energy term \cite{geshkenbein_vortices_1987}. 
An odd number of the $\pi$ phase shifts accumulated around a superconducting loop will reverse minima and maxima of the total free energy . 
It is thus the maximum of $T_c$, instead of the minimum, that occurs when $n-\Phi/\Phi_0=\pm\frac{1}{2}$, as shown in Fig.~1B. 
The fluxoid quantization becomes $\Phi^{\prime} = (n+1/2)\Phi_0$. 
In the absence of an external magnetic field, such a superconducting loop will hold a HQF as its ground state. 
GLB concluded that such a circumstance is most likely to occur in polycrystalline $p-$wave SCs \cite{geshkenbein_vortices_1987}. 

To examine the fluxoid quantization in $\beta$-Bi$_2$Pd, we fabricated sub-$\mu$m-sized ring devices using 50 nm-thick (001)-textured $\beta$-Bi$_2$Pd thin films, deposited on oxidized silicon substrates by magnetron sputtering (Section I of \cite{SM}). 
The size of the ring is important because the oscillation occurs in units of $\Phi_0 \approx$ 20~Oe-($\mu$m)$^2$. 
For a ring of 1~$\mu$m $\times$ 1~$\mu$m, oscilaltion occurs in field increment of about 20 Oe. 
Measuring much larger rings is more demanding on the field resolution, and creates difficulty determining the zero-exteranl-field state in the presence of the remnant field. 
For smaller rings, the size becomes comparable to the coherent length.
A representative device geometry can be found in Fig.~1C with a mean size of 0.8~$\mu$m $\times$ 0.8~$\mu$m. 
Control samples with the same device geometries were patterned on 28 nm-thick thin films of Nb, which is a conventional $s-$wave SC. 
The temperature dependence of the device resistance is shown in Fig.~1D. 
A broadening of the transition width ($\sim$ 1.5~K) as compared to that of the as-grown films ($<$ 0.2~K) is typical for devices in similar dimensions that have undergone nanofabrication processes \cite{Gammel1990,Sochnikov2010,Carillo2010,cai_unconventional_2013,yasui_little-parks_2017}. 
The Little-Parks effect can be observed when the sample is placed at a fixed temperature within the superconducting transition regime, where the variation of the $T_c$ manifests as a fluctuation of the resistance \cite{Little1962}. 
Fig.~1E presents a typical result of the Little-Parks experiment, obtained from the Nb ring device. 
The observed oscillation period is 30.2 Oe, in good agreement with 32.3 Oe as expected for $\Phi_0$ of the 800~nm $\times$ 800~nm ring area. 
Importantly, the resistance minima, which correspond to the $T_c$ maxima, occur when $\Phi=n\Phi_0$, as routinely observed for $s-$wave SCs \cite{Parks1964} and high-$T_c$ cuprate $d$-wave SCs \cite{Gammel1990,Carillo2010}, both singlet SCs.  
Up to about 70 oscillations have been observed on a roughly parabolic-shaped background. 
The aperiodic background is commonly observed in Little-Parks experiments, believed to originate from the misalignment of the magnetic field and the finite line-width of the ring \cite{Tinkham1963,Parks1964,Moshchalkov1995}. 
As a precaution against trapping vortices, the measurements were always performed after zero-field cooling from 10~K. 

The Little-Parks oscillation of the $\beta$-Bi$_2$Pd ring device is shown in Fig.~2A. 
The same oscillation period of $\sim$ 30~Oe is observed. 
The aperiodic background can be subtracted from the raw data, and the $\Delta R$ versus $H$ oscillation is presented in Fig.~2B. 
The magnitude of the resistance oscillation translates to $\sim$ 0.015~K variations of $T_c$, consistent with theoretical expectations for the Little-Parks effect \cite{TINKHAM1964}. 
In stark contrast with the case of Nb, however, $\Phi=n\Phi_0$ now corresponds to the resistance \textit{maxima} (or the $T_c$ minima). 
The resistance minima and the $T_c$ maxima are instead observed when $\Phi=(n+1/2)\Phi_0$. 
The $\frac{1}{2}\Phi_0$ shift of the free energy minima indicates that it is energetically preferred that fluxoid quantization takes $\Phi^{\prime} = (n+1/2)\Phi_0$. 
When the external magnetic field is zero, the superconducting ring circulates a finite $\vec{j_s}$ to sustain one HQF, which accounts for the lowest $T_c$ and the highest free energy. 
The system is more energetically satisfied when the external field supplies $\pm\frac{1}{2}\Phi_0$, whereas $\vec{j_s}$ can rest to zero. 
The symmetric shape of the magnetoresistance oscillation trace shows that the effect is not due to defect-trapped vortices. 
We also verify that the result is robust against the different field sweeping directions and current densities (Sections III and IV, Fig S3 of \cite{SM}). 
The HQF is repeatedly observed in numerous $\beta$-Bi$_2$Pd rings with various geometrical device shape factors (Section II of \cite{SM}). 

In Fig.~3 we demonstrate the temperature dependence of the Little-Parks effect in $\beta$-Bi$_2$Pd. 
The $\pi$ phase shift persists from the emergence of the Little-Parks effect at 2~K, through the highest temperature of 2.6~K, before the oscillation disappears presumably due to the loss of coherence over the length scale of the ring device. 
The lack of temperature variation suggests the predominance of the unique pairing symmetry that gives rise to the HQF, without any detectable dilution of possible $s$-wave components \cite{kirtley_temperature_1999}. 

In the following we discuss the implication of our experimental observations. 
The GLB HQF is an experimental signature for unconventional pairing mechanisms, not restricted to the $p-$wave pairing. 
In fact, HQF was first observed in $d-$wave SC tricrystals \cite{tsuei_pairing_1994}. 
The orientations of the crystal boundaries had to be carefully designed; otherwise HQF would not exist \cite{tsuei_pairing_1994,Kirtley1995}.
To our best knowledge, HQF has not been observed in polycrystalline specimens of high-$T_c$ SCs, although the Little-Parks experiment has been conducted on microstructured mesh of polycrystalline YBa$_2$Cu$_3$O$_7$ thin films \cite{Gammel1990}. 
In our experiment, on the other hand, HQFs are observed in the overwhelming majority of the $\beta$-Bi$_2$Pd ring devices (Sections II and V of \cite{SM}). 
Considering the previous literature reporting the link between $\beta$-Bi$_2$Pd and the spin-triplet pairing \cite{Sakano2015,Iwaya2017,Lv2017}, a contributing factor to the robustness of the HQF state might be the unique anisotropy of the $p-$wave pairing. 
The order parameter reverses sign when rotating 180$^{\circ}$, whereas for $d-$wave the sign reversal occurs upon rotating 90$^{\circ}$. 
This protects the HQF from moderate disorders on the crystal grain boundaries, comparing to that gauged for $d$-wave SCs \cite{tsuei_pairing_1994,Kirtley1995}. 
Nevertheless, the observation of HQF is an unequivocal evidence for unconventional superconductivity, consistent with previous experiments concluding topological superconductivity in $\beta$-Bi$_2$Pd \cite{Sakano2015,Iwaya2017,Lv2017}. 
Although our Little-Parks experiment cannot decisively distinguish among different non-$s$-wave pairing mechanisms, 
the Andreev-reflection spectroscopy measurements on the $\beta$-Bi$_2$Pd thin films conclusively show that it is a spin-triplet SC with $p$-wave pairing. 
This result will be presented elsewhere. 

To conclude, we observed HQFs in superconducting $\beta$-Bi$_2$Pd thin films, evidenced by the $\pi$ phase shift of the Little-Parks oscillations. 
The presence of HQFs unambiguously indicates unconventional pairing mechanism for superconductivity. 
Together with tunneling spectroscopy \cite{Iwaya2017,Lv2017} and angle-resolved photoemission spectroscopy results \cite{Sakano2015}, the $p-$wave spin-triplet pairing can be concluded. 
Our observation strongly supports $\beta$-Bi$_2$Pd as a topological superconductor, which would be a suitable candidate material, an intrinsic TSC in which to search for Majorana fermions. 
The confirmation of the spin-triplet SC may open other venues for condensed matter physics.  




\bibliographystyle{Science}

\section*{Acknowledgments}
	This work was supported by the U.S. Department of Energy, Basic Energy Science, Award Grant No. DESC0009390. We thank Yi Li and C.C. Tsuei for valuable discussions. The e-beam lithography was conducted at the UNDF lab of University of Delaware and the NanoFab lab of NIST (CNST). We thank Kevin Lister and Richard Kasica for assistance in the nanofabrication processes. 



\clearpage


\begin{figure}
	\centering
	\includegraphics[width=16cm]{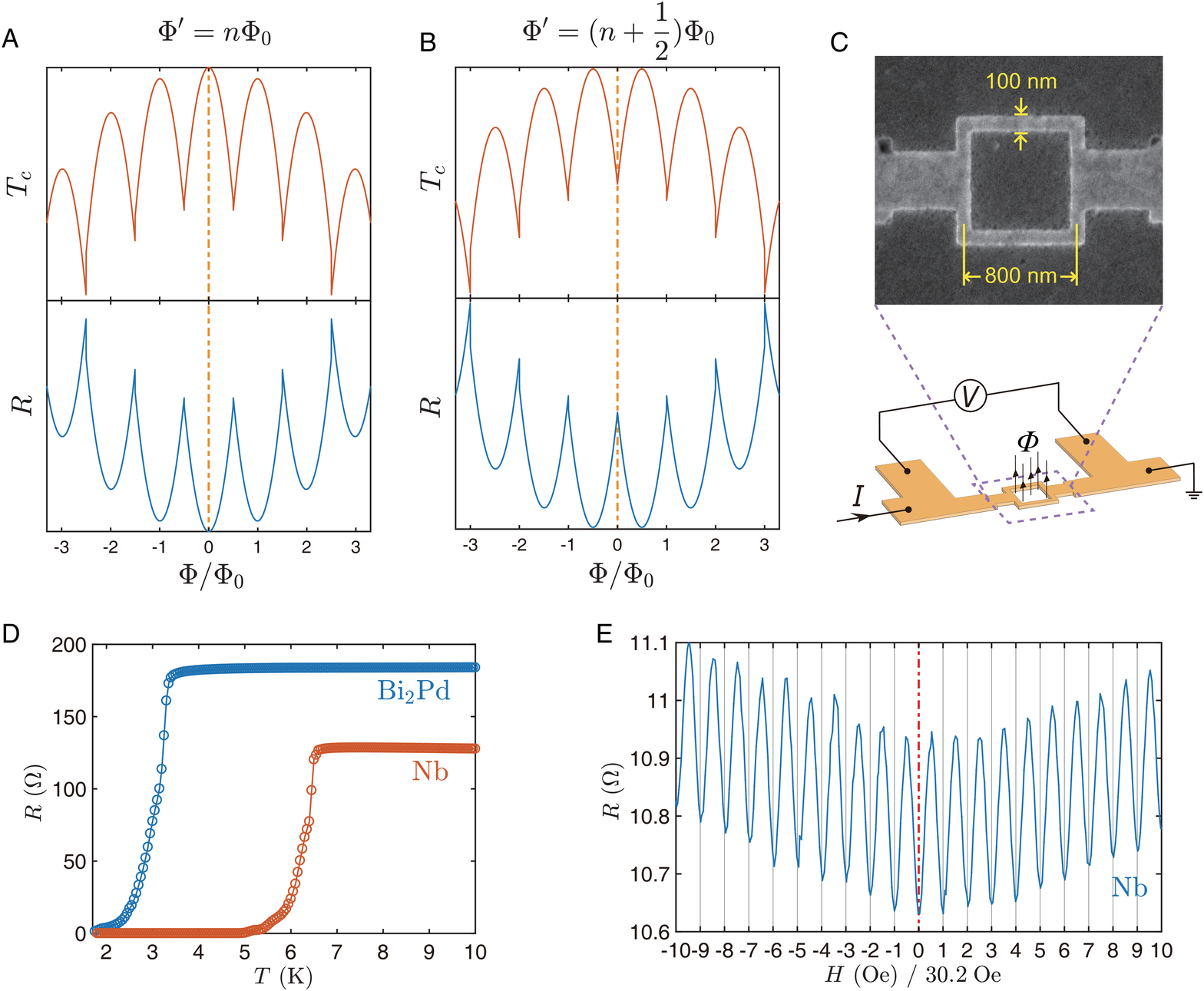}
	\caption{\label{fig1}
		\textbf{Little-Parks effect and the experimental setup.} 
		\textbf{(A)} Schematic drawing of the Little-Parks effect of conventional $s-$wave SC. 
		The critical temperature (upper panel) and the resistance (lower panel) oscillate as a function of the applied magnetic flux, in the period of one magnetic flux quantum $\Phi_0 = hc / 2e$. 
		The fluxoid takes integer-quantized values of $\Phi^{\prime}=n\Phi_0$. 
		\textbf{(B)} Schematic drawing of the effect of the half-quantum fluxoid manifested as the $\pi$ phase shift in the Little-Parks oscillation. 
		The fluxoid takes fractionally quantized values of $\Phi^{\prime}=(n+1/2)\Phi_0$. 
		\textbf{(C)} The scanning electron microscope image of a representative superconducting ring device (upper panel) and the schematic drawing of the experimental setup (lower panel). 
		\textbf{(D)} Temperature dependence of the resistance of the $\beta$-Bi$_2$Pd and Nb ring devices. 
		The film thicknesses of $\beta$-Bi$_2$Pd and Nb are 50~nm and 28~nm, respectively. 
		\textbf{(E)} The Little-Parks effect of the Nb ring device, showing the ordinary case depicted in (A). 
		The resistance oscillates as a function of the perpendicular magnetic field. 
		The sample was held at a constant temperature of 5.7~K, in the vicinity of the normal-superconducting phase transition. 
	}
\end{figure}

\begin{figure}
	\centering
	\includegraphics[width=16cm]{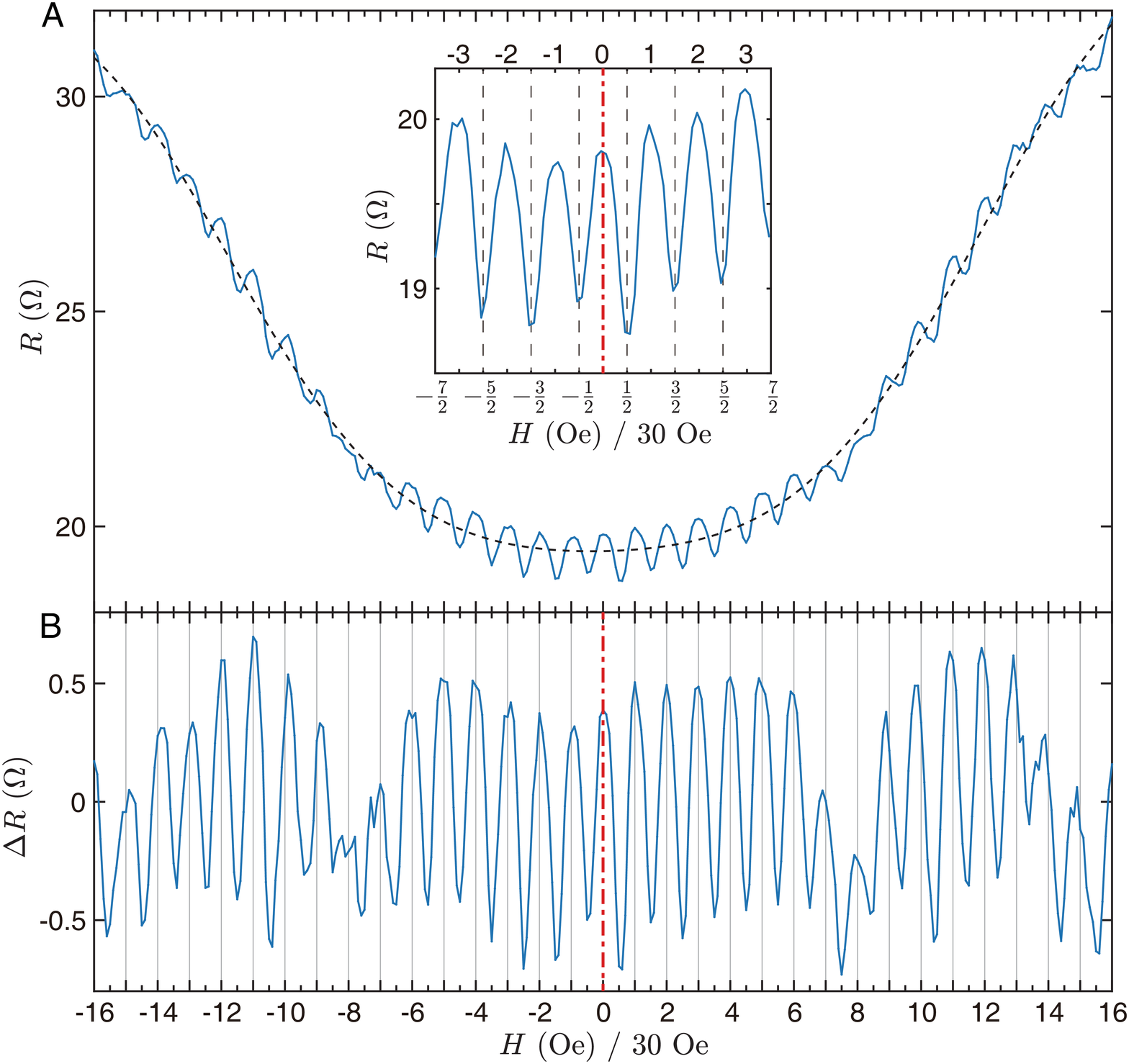}
	\caption{\label{fig2}
		\textbf{Little-Parks effect of the $\beta$-Bi$_2$Pd ring device.} 
		\textbf{(A)} The oscillation of the $\beta$-Bi$_2$Pd ring device resistance as a function of the perpendicularly applied magnetic field. 
		The sample was held at a constant temperature of 2.5~K. 
		The x-axis is displayed in the unit of the oscillation period of 30~Oe, in agreement with the expected magnetic flux quantum for the device geometry. 
		The black dashed line denotes the aperiodic background. 
		The inset shows the zoom-in view of the low field region of (A). 
		The vertical black dashed lines denote the applied magnetic flux of $\Phi=(n+1/2)\Phi_0$, which correspond to the oscillation minima. 
		The red dot-dashed line denotes the zero external field.  
		\textbf{(B)} The Little-Parks oscillation in which the aperiodic background has been subtracted from the raw data as presented in (A). 
		The gray vertical lines denote the the applied magnetic flux of $\Phi=n\Phi_0$, which correspond to the oscillation maxima. 
	}
\end{figure}

\begin{figure}
	\centering
	\includegraphics[width=12cm]{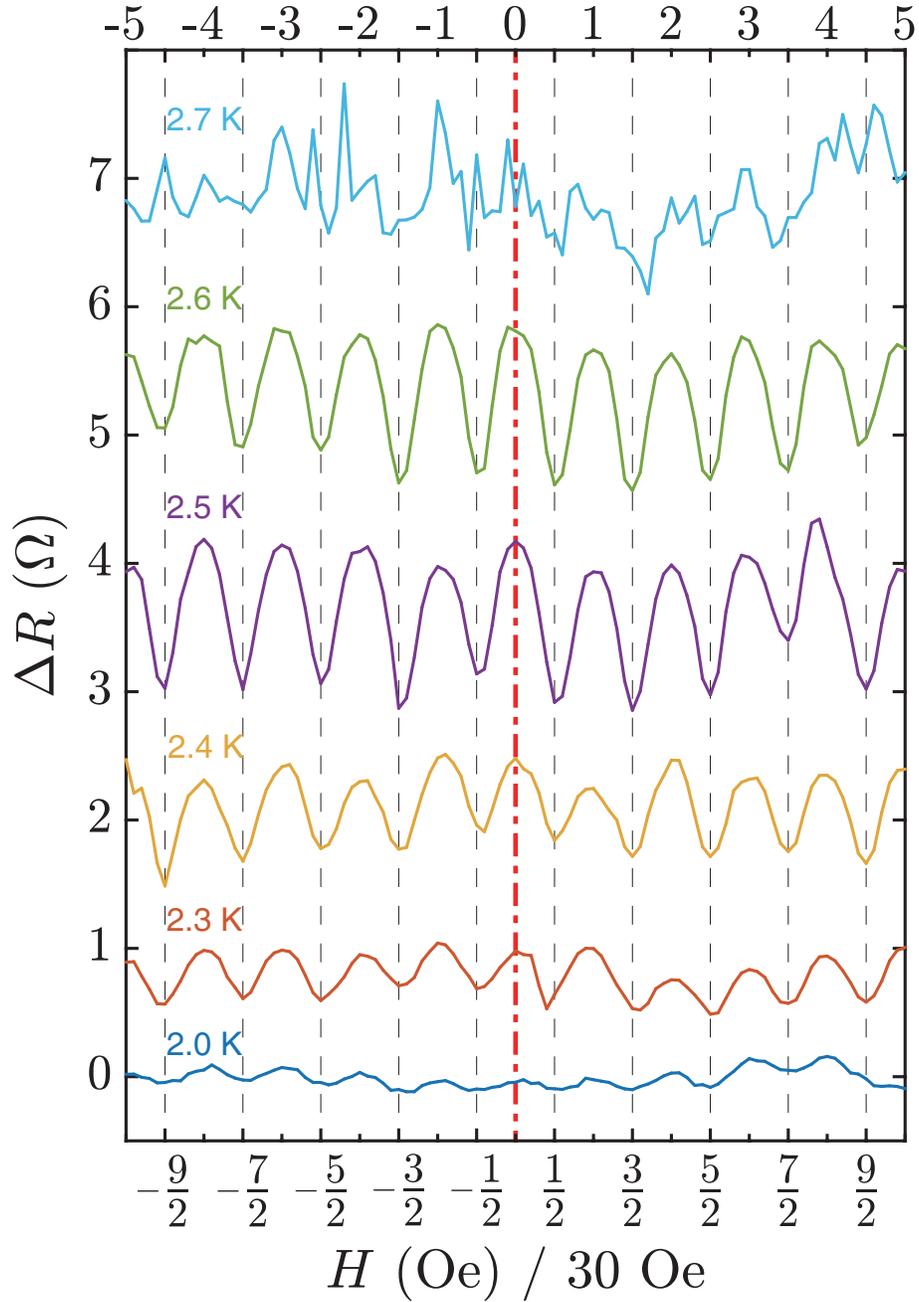}
	\caption{\label{fig3}
		\textbf{Temperature evolution of the Little-Parks effect of the $\beta$-Bi$_2$Pd ring device.} 
		The oscillation of the resistance as a function of applied magnetic field with the aperiodic background subtracted. 
		The vertical black dashed lines denote the applied magnetic flux of $\Phi=(n+1/2)\Phi_0$. 
	}
\end{figure}

\end{document}